\documentclass[twocolumn,showpacs,preprintnumbers,amsmath,amssymb]{revtex4}

\usepackage{graphicx}
\usepackage{dcolumn}
\usepackage{bm}

\begin{document}

\title{Equilibrium long-ranged charge correlations at the surface \\
of a conductor coupled to the electromagnetic radiation II}
\author{Bernard Jancovici}
\email{Bernard.Jancovici@th.u-psud.fr}
\author{Ladislav \v{S}amaj}
\altaffiliation[On leave from ]
{Institute of Physics, Slovak Academy of Sciences,
Bratislava}
\email{Ladislav.Samaj@savba.sk}
\affiliation{Laboratoire de Physique Th\'eorique,
Universit\'e de Paris-Sud \\
91405 Orsay Cedex, France\footnote{Unit\'e Mixte de Recherche No 8627-CNRS}}

\date{\today}

\begin{abstract}
Results of a previous article with the same title are retrieved by a different
method. A one-component plasma is bounded by a plane surface. The plasma is
fully coupled to the electromagnetic field, therefore the charge correlations
are retarded. The quantum correlation function of the surface charge densities,
at times different by $t$, at asymptotical large distances $R$, at inverse
temperature $\beta$, decays as $-1/(8\pi^2\beta R^3)$, a surprisingly simple
result: the decay is independent of Planck's constant $\hbar$ and of the
time difference $t$. The present paper is based on the analysis of the
collective vibration modes of the system.
\end{abstract}


\maketitle

\section{Introduction}
In a previous paper \cite{SJ}, we studied the asymptotic form of the 
two-point correlation function of the surface charge densities on 
a plane wall bounding a conductor [in the special case when the conductor 
is a one-component plasma (OCP), also called jellium], taking into account 
the retardation and the quantum nature of both the one-component plasma 
and the radiation. 
The novelty was the retardation; instead of assuming as an interaction 
the Coulomb potential only, in \cite{SJ} the coupling was through
the full electromagnetic radiation. 
In other words instead of assuming the velocity of light $c$ to be infinite, 
the full Maxwell equations were used.

The previous paper \cite{SJ} used the elaborate formalism of Rytov \cite{R}, 
presented also in \cite{LP}. 
This formalism is macroscopic, using frequency-dependent dielectric functions. 
In the present paper, we retrieve the same results by a simpler method, 
based on the analysis of the collective vibration modes of the system. 
This method is partially microscopic. 
It has already been used in the non-retarded case \cite{J}, Sec. 4.

We use Gaussian units.
The OCP is made of point-particles of charge $e$, mass $m$, and number density
$n$, immersed in a uniform neutralizing background of charge density $-ne$. 
We recall the geometry (Fig. 1). 
We use Cartesian coordinates; a point is ${\bf r}=(x,y,z)$. 
The OCP occupies the half-space $\Lambda_1=\{x>0\}$, the half-space 
$\Lambda_2=\{x<0\}$ is vacuum; the two half-spaces are separated by
an plane wall, impenetrable to the jellium, at $x=0$. 
A point on the wall is ${\bf R}=(y,z)$.

\begin{figure}[b]
\begin{center}
\includegraphics[width=0.45\textwidth,clip]{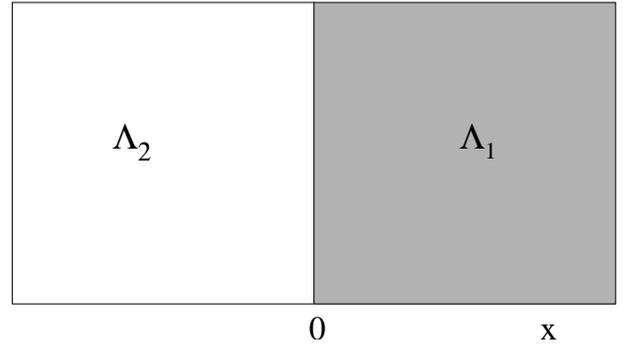}
\caption{The geometry.}
\end{center}
\end{figure}

After long calculations, a very simple result was found in \cite{SJ}: 
\begin{eqnarray}
\beta S(t,{\bf R}) & \equiv &
\beta\frac{1}{2}\langle\sigma(t,{\bf R})\sigma(0,{\bf 0})+\sigma(0,{\bf 0}) 
\sigma(t,{\bf R})\rangle^{\rm T} \nonumber \\
& \sim & -\frac{1}{8\pi^2}\frac{1}{R^3} ,\qquad R\to\infty , \label{1}
\end{eqnarray}
where $\beta$ is the inverse temperature, $\langle\cdots\rangle^{\rm T}$
represents a truncated statistical average, $\sigma(t,{\bf R})$ is the surface
charge density at time $t$ and at point ${\bf R}$ on the surface. 
This value (\ref{1}) surprisingly is independent of $t$ and $\hbar$, 
boiling down to the classical result at time difference zero without 
retardation.

We define the Fourier transform of a function $f({\bf R})$ as
\begin{equation} \label{2}
f({\bf q})=\int d^2 R\, \exp(i{\bf q}\cdot{\bf R})f({\bf R}).
\end{equation}
A result equivalent to (\ref{1}) is that the Fourier transform of its lhs,
$\beta S(t,{\bf q})$, has a kink singularity at ${\bf q}={\bf 0}$, behaving at
small $q$ like $q/(4\pi)$. 
 
The present paper is organized as follows. Sec. II is a general exposition of
the formalism of vibration modes. Sec. III describes the contribution of
surface modes (which are localized on both sides of the wall). Sec. IV
describes the contribution of transverse modes delocalized on the vacuum
side. Sec. V describes the contribution of transverse modes delocalized on
both sides. In Sec. VI, we recall the contribution of the longitudinal
modes. Sec. VII is the Conclusion. 
 
\section{Collective vibration modes}
We consider a fluctuation of the surface charge density $\sigma$, of
wavenumber ${\bf q}$ and frequency $\omega$, such that $\sigma$ is of the form
\begin{equation}
\sigma({\bf R},t)= \sigma_{{\bf q}\omega}(t)\exp(i{\bf q\cdot R})+{\rm c.c.}\, , 
\label{3}
\end{equation}
where c.c. means complex conjugate. $\sigma_{{\bf q}\omega}(t)$ is a complex
quantity, vibrating at frequency $\omega$. 
The emitted radiation, of frequency $\omega$, is described by the
Maxwell equations (we use the microscopic ones, involving only the electric
field ${\bf E}$ and the magnetic field ${\bf B}$, averaged in a suitable way
\cite{Jackson}). 
The charge density is $\rho$, the electric current density is ${\bf J}$, 
the velocity of light is $c$. 
These Maxwell equations are  
\begin{eqnarray}
\nabla\times{\bf B} & = & \frac{1}{c}\frac{\partial{\bf E}}{\partial t}
+\frac{4\pi}{c}{\bf J}, \label{4} \\ 
\nabla\times{\bf E} & = & -\frac{1}{c}\frac{\partial{\bf B}}{\partial t},
\label{5} \\
\nabla\cdot{\bf E} & = & 4\pi\rho, \label{6} \\
\nabla\cdot{\bf B} & = & 0 . \label{7}
\end{eqnarray} 
In the quantum case, the quantities appearing in these equations are
operators. 
In region $\Lambda_2$, $\rho=0$ and  ${\bf J}={\bf 0}$, of course. 

There are solutions to these Maxwell equations which are superpositions of
transverse waves, which we shall study first. 
In region $\Lambda_1$, for these transverse waves, $\rho=0$. 
Furthermore, the Maxwell equations (\ref{4}-\ref{7}) have to be 
supplemented by  
\begin{equation} \label{8}
{\bf E} = \frac{4\pi}{\omega_p^2}\frac{\partial{\bf J}}{\partial t},
\end{equation}
where $\omega_p=(4\pi ne^2/m)^{1/2}$ is the plasma frequency. 
This equation (\ref{8}) is obtained from assuming that the OCP can be 
treated in a hydrodynamical linearized approximation where the velocity is 
${\bf v}({\bf r},t)$ obeying the Newton equation 
$m\partial{\bf v}/\partial t=e{\bf E}$ (the term $e{\bf v}\times{\bf B}$ in
the Lorentz force is suppressed by the linearization) and the current density 
is ${\bf J}=en{\bf v}$ (the density is the constant $n$, again because of the
linearization). 
It should be remarked that in the equation (\ref{8}), there is
no damping term; this absence of damping term is valid for the small
wavenumbers which will be considered, and is a property special to the OCP.

Macroscopic Maxwell equations equivalent to (\ref{4}-\ref{8}) can also be 
obtained. 
Combining (\ref{4}) and (\ref{8}), at frequency $\omega$, we obtain
\begin{equation} \label{9}
\nabla\times{\bf B}=\frac{1}{c}\frac{\partial(\epsilon{\bf E})}{\partial t},
\end{equation}
where the frequency-dependent dielectric function is
\begin{equation} \label{10}
\epsilon(\omega)=1-\frac{\omega_p^2}{\omega^2},
\end{equation}
which is of the Drude form, \emph{without dissipation}. 
Multiplying (\ref{6}), where $\rho=0$, by $\epsilon$ does not change anything 
if $\epsilon\neq 0$ (the case $\epsilon=0$ will be studied later). 
To an excellent approximation, the magnetic permeability is $\mu=1$. 
Thus we have obtained the macroscopic Maxwell equations \cite{Jackson}, 
with now $\rho=0$ and ${\bf J}=0$.

For these transverse waves, the wavenumber vector has components 
$(k_1,{\bf q})$ in region $\Lambda_1$ and $(k_2,{\bf q})$ in region
$\Lambda_2$, where $k_1$ and $k_2$ are the $x$ components (the wavenumbers
have the same components parallel to the surface ${\bf q}=(q_y,q_z)$ as a
consequence of the boundary conditions, as will be shown later).

From the Maxwell equations (\ref{4}-\ref{7}) and (\ref{8}), the dispersion
relations are  
\begin{equation} \label{11}
\omega^2=c^2(q^2+k_2^2)
\end{equation}
in region $\Lambda_2$, and 
\begin{equation} \label{12}
\omega^2=\omega_p^2+c^2(q^2+k_1^2)
\end{equation}
in region $\Lambda_1$.  
Depending on the value of $\omega$, $k_1$ or $k_2$ are real or pure imaginary, 
as shown from (\ref{11}) and (\ref{12}). 
If $\omega^2<(cq)^2$, both $k_1$ and $k_2$ are pure imaginary. 
If $(cq)^2<\omega^2<\omega_p^2+(cq)^2$, $k_1$ is pure imaginary and $k_2$ is
real. If $\omega^2>\omega_p^2+(cq)^2$, both  $k_1$ and $k_2$ are real.
 
In addition to these plane transverse waves, in region $\Lambda_1$, when the
frequency is $\omega_p$ (then $\epsilon=0$) the Maxwell equations have
solutions which are longitudinal waves \cite{Jackson,LL}. 
They have already been studied in \cite{J}.

The energy density in region $\Lambda_2$ is \cite{Jackson}
\begin{equation} \label{13}
U_2=\frac{1}{8\pi}(E^2+B^2);
\end{equation} 
the energy density in region $\Lambda_1$ is 
\begin{equation} \label{14}
U_1=\frac{1}{8\pi}(E^2+B^2)+\frac{2\pi}{\omega_p^2}J^2
\end{equation}
A derivation of (\ref{14}) is presented in the Appendix, as well as another
form of it \cite{LL}.
 
At $x=0$, the existence of a surface charge density $\sigma$ is associated
with a discontinuity of the $x$ component of the electrical field
\begin{equation} \label{15}
4\pi\sigma=E_x^+ -E_x^-,
\end{equation}
where the upperscripts $+$ or $-$ mean approaching the plane $x=0$ from the
regions $\Lambda_1$ or $\Lambda_2$, respectively.
The other conditions at $x=0$ are the continuity of the components of 
the electric and magnetic fields parallel to the surface. 
From the conditions at $x=0$, one easily deduces that the $(y,z)$ components 
of the wavenumbers indeed are ${\bf q}$. 
Since the system is invariant by rotations around the $x$ axis,
general results can be obtained by choosing ${\bf q}$ along the $y$ axis:
${\bf q}=(q_y,0)$; then $q_y^2=q^2$. 
We consider only modes which contribute to $\sigma$.

\section{$\omega^2<(cq)^2$ (polaritons)}
Since $k_1$ and $k_2$ are pure imaginary, the solutions of the Maxwell
equations are of the form (the subscript $j=1,2$ denotes the region 
$\Lambda_1$ or $\Lambda_2$) \cite{P}
\begin{equation} \label{16}
{\bf E}_j = [a_j(t),b_j(t),0]\exp(i q_y y-\kappa_j\vert x\vert )+{\rm c.c.}\, ,
\end{equation}
\begin{equation} \label{17} 
{\bf B}_j =  [0,0,d_j(t)]\exp(i q_y y -\kappa_j \vert x\vert )+{\rm c.c.}\, ,  
\end{equation}
where, from (\ref{11}) and (\ref{12}), $c\kappa_2=\sqrt{(cq)^2-\omega^2}$ and
$c\kappa_1=\sqrt{\omega_p^2+(cq)^2-\omega^2}$. 
Since all fields are localized near the surface, these modes are called 
surface plasmons; polariton is also used when retardation, as here, 
is taken into account.

From the Maxwell equations and the conditions that $E_y$ and 
$B_z$ are continuous at $x=0$, in this Section there is a relation between 
$\omega$ and ${\bf q}$:  
\begin{equation} \label{18}
\omega^2=\omega_p^2/2 +(cq)^2-\sqrt{(\omega_p^2/2)^2+(cq)^4}. 
\end{equation}
Therefore, in this Section, $\sigma_{{\bf q}\omega}$ will be called 
$\sigma_{\bf q}$.  
$\omega$ has its non-retarded value $\omega_p/\sqrt{2}$ only at large $q$. 
For the small values of $q$, that we are interested in, $\omega$
behaves as $cq$. 
From (\ref{18}) follows $\kappa_1\kappa_2=q^2$, a useful relation in 
the calculations (the detail of which is omitted) which follow.

Taking into account (\ref{15}) and the continuity of $E_y$ and $B_z$, the
Maxwell equations (\ref{5}), (\ref{6}) [$\rho=0$ in (\ref{6})] give the
prefactors $a_i,b_i,d_i$ as functions of $\sigma_{{\bf q}}$. 
One finds
\begin{eqnarray}
& & a_1(t)=\frac{\kappa_2}{\kappa_1+\kappa_2}4\pi\sigma_{\bf q}(t) ,\quad
a_2(t)=-\frac{\kappa_1}{\kappa_1+\kappa_2}4\pi\sigma_{\bf q}(t), \nonumber \\
& & b_1(t)=b_2(t)=- i \frac{q_y}{\kappa_1+\kappa_2}4\pi\sigma_{\bf q}(t),
\label{19} \\
& & d_1(t)=d_2(t)=\frac{iq_y}{c\kappa_2}\frac{1}{\kappa_1+\kappa_2}
4\pi\dot{\sigma}_{\bf q}(t) . \nonumber 
\end{eqnarray}
In region $\Lambda_1$, ${\bf J}$ can be obtained from (\ref{4}). One finds
\begin{eqnarray}
& & J_{1x}=-\exp(iq_y y-\kappa_1 x)\dot{\sigma}_{\bf q}(t)+{\rm c.c.}, 
\nonumber \\
& & J_{1y}=\frac{iq_y}{\kappa_2}\exp(iq_y y-\kappa_1 x)
\dot{\sigma}_{\bf q}(t)+{\rm c.c.}. \label{20} 
\end{eqnarray}
From (\ref{19}) and (\ref{20}), one computes the energy densities
(\ref{13}) and (\ref{14}) and the total energy $H_{\bf q}$. 
Later, we shall consider that the large area $A$ of the wall goes to infinity; 
therefore, the oscillatory terms $\exp(\pm 2 i q_yy)$ do not contribute to 
the integral on ${\bf R}$.
One finds 
\begin{eqnarray}
H_{\bf q} & \equiv & \int_A d^2 R\left[\int_0^{\infty}d x\, U_1+
\int_{-\infty}^0 d x\, U_2\right]  \nonumber \\ & = &
A\, 2\pi \left( \frac{\kappa_1}{\kappa_2}+\frac{\kappa_2}{\kappa_1} \right)
\frac{1}{\kappa_1+\kappa_2}\nonumber \\
& & \times \left[\vert \sigma_{\bf q}(t)\vert^2+\frac{1}{\omega^2}
\vert\dot{\sigma}_{\bf q}(t)\vert^2\right]  \label{21} \\
& \equiv & AC_ {\bf q}\left[\vert \sigma_{\bf q}(t)\vert^2+\frac{1}{\omega^2}
\vert\dot{\sigma}_{\bf q}(t)\vert^2\right] \nonumber 
\end{eqnarray}
One sees that $H_{\bf q}$ is the energy of a two-dimensional harmonic
oscillator (two-dimensional because $\sigma_{\bf q}(t)$ is a complex
quantity). For such a quantum oscillator, where the variable 
$\sigma_{\bf q}(t)$ plays the role of the position variable, the contribution 
to $\beta S(t,{\bf q})$ is 
\begin{equation} \label{22}
\beta S_A(t,{\bf q})=\frac{1}{C_{\bf q}}f(\omega)\cos(\omega t) , 
\end{equation}
where
\begin{equation} \label{23}
f(\omega)=\frac{\beta\hbar\omega}{2}\coth\frac{\beta\hbar\omega}{2}.
\end{equation}
Here, for $q\rightarrow 0$, $\omega\sim cq$, and (\ref{22}) behaves like
$q^2$. Therefore, the polaritons do not contribute to the retarded asymptotic 
form of $S(t,{\bf R})$.

On the contrary, in the non-retarded case $c\rightarrow\infty$,
$\omega=\omega_s\equiv\omega_p/\sqrt 2$, and (\ref{22}) becomes
\begin{equation} \label{24}
\beta S_A(t,q)=\frac{q}{2\pi}f(\omega_s)\cos(\omega_s t), 
\end{equation}
in agreement with \cite{J}.

\section{$(cq)^2<\omega^2<\omega_p^2+(cq)^2$}
Now $k_1$ is pure imaginary while $k_2$ is real. With ${\bf q}=(q_y,0)$, in
region $\Lambda_2$ the general form of component $E_x$ must be 
\begin{equation} \label{25}
E_{2x}=\exp( i q_y y)[a(t)\cos(k_2x)+b(t)\sin(k_2x)]+
{\rm c.c.}\, . 
\end{equation}
Using the Maxwell equations (\ref{5}), (\ref{6}) gives the other non-zero
components of the fields in region $\Lambda_2$ as
\begin{equation} \label{26}
E_{2y}=\exp( i q_y y)\frac{ i k_2}{q_y}   
[-a(t)\sin(k_2x)+b(t)\cos(k_2x)]+{\rm c.c.}\, , 
\end{equation}
\begin{equation} \label{27}
B_{2z}=-\exp( i q_y y)\frac{1}{cq_y}
[\dot{a}(t)\cos(k_2x)+\dot{b}(t)\sin(k_2x)]+{\rm c.c.}\, . 
\end{equation}
In region $\Lambda_1$, since $k_1$ is pure imaginary, the fields 
depend on $x$ like $\exp(-\kappa_1x)$. 
The continuity of $E_y$ and $B_z$ at $x=0$ determine the coefficients 
in function of $a$ and $b$, and (\ref{3}) and (\ref{15}) give
\begin{equation} \label{28}
a(t)=\frac{\omega^2-\omega_p^2}{\omega_p^2}4\pi\sigma_{{\bf q}\omega}(t),\qquad
b(t)=-\frac{\kappa_1\omega^2}{k_2\omega_p^2}4\pi\sigma_{{\bf q}\omega}(t). 
\end{equation}

Let $\Lambda_2$ be the region $0>x>-L_2$ (with a large $L_2$, which at the end 
will be taken as infinite). 
Since $U_1$ has an exponential factor $\exp(-2\kappa_1x)$, the region 
$\Lambda_1$ does not contribute to the total energy in this limit. 
The total energy only is $H_{{\bf q}\omega} = A \int_{-L_2}^0 d x\, U_2$. 
Using (\ref{25}-\ref{28}) in (\ref{13}) gives 
\begin{eqnarray}
&&H_{{\bf q}\omega}=2\pi A L_2\frac{\omega^2(\omega_+^2-\omega^2)
(\omega^2-\omega_-^2)}{\omega_p^2(cq)^2[\omega^2-(cq)^2]} \nonumber \\
&&\times\left[\vert\sigma_{{\bf q}\omega}(t)\vert^2+\frac{1}{\omega^2}
\vert\dot{\sigma}_{{\bf q}\omega}(t)\vert^2\right]  \nonumber \\
&&\equiv A L_2 C_{{\bf q}\omega}\left[\vert \sigma_{{\bf q}\omega}(t)\vert^2
+\frac{1}{\omega^2}\vert\dot{\sigma}_{{\bf q}\omega}(t)\vert^2\right],  
\label{29}
\end{eqnarray}
where $\omega_-^2$ is given by (\ref{18}) and $\omega_+^2$ is given by a
modified (\ref{18}) with a $+$ sign in front of the square root. 
Therefore, the contribution of this mode to $\beta S(t,{\bf q})$ is 
\begin{equation} \label{30}
\beta S_{B\omega}(t,{\bf q})=\frac{1}{L_2C_{{\bf q}\omega}}
f(\omega)\cos(\omega t). 
\end{equation}

There are an infinity of modes of this kind, labeled by their frequency 
$\omega$, or, equivalently, by $k_2$. 
For obtaining the total contribution of these modes, 
$\beta S_B(t,{\bf q})$ to $\beta S(t,{\bf q})$, one has to take
the sum on $k_2$ of (\ref{30}).
Since $L_2$ is large, $\sum_{k_2}\cdots=(L_2/\pi)\int d k_2\cdots$. 
Furthermore, $d k_2= d (\omega^2)/(2c^2k_2)$. 
Therefore,
\begin{eqnarray}
\beta S_B(t,{\bf q}) & = &\frac{1}{\pi}\int_{(cq)^2}^{\omega_p^2+(cq)^2}
d(\omega^2)\, \frac{1}{2c{\sqrt{\omega^2-(cq)^2}\, C_{{\bf q}\omega}}}
\nonumber \\ & & \times f(\omega)\cos(\omega t). \label{31}
\end{eqnarray}
After the change of variable $\omega^2=(cq)^2(1+u)$, (\ref{31}) becomes, in
the limit $q\rightarrow 0$,
\begin{equation} \label{32}
\beta S_B(t,{\bf q})\sim\frac{q}{4\pi^2}f(0)\cos0
\int_0^{\infty}\frac{d u} {\sqrt{u}(1+u)}=\frac{q}{4\pi}.
\end{equation}
(\ref{32}) will be found the only contribution of order $q$ to 
$\beta S(t,{\bf q})$.

\section{$\omega^2>\omega_p^2+(cq)^2$}
Now, $k_1$ and $k_2$ are both real. The general form of the components $x$ of
the electrical fields are
\begin{equation}
E_{jx}=\exp( i q_y y)[a_j(t)\cos(k_j x)+b_j(t)\sin(k_j x)]+
{\rm c.c.}\, . \label{33}
\end{equation}
(\ref{33}) involves 4 coefficients, instead of 3 in $E_{ix}$ of Section IV
(where $E_{1x}$, not written explicitly, involves only 1 coefficient). 
Therefore the equations which were used in Section IV are not enough for 
determining all the coefficients in the fields as functions of 
$\sigma_{{\bf q}\omega}(t)$. 
Fortunately, parts of the fields are uncoupled to $\sigma$ and, 
for describing the modes coupled to $\sigma$, it is enough to choose in 
(\ref{33}) $b_i=0$. 
Then, the Maxwell equations and the conditions at $x=0$ determine
all the fields as functions of $\sigma_{{\bf q}\omega}(t)$. 
Assuming region $\Lambda_i$ to be of large length $L_i$ in $x$, 
one finds the total energy $H=H_{1{\bf q}\omega}+H_{2{\bf q}\omega}$ 
where the energy in region $\Lambda_1$ is
\begin{eqnarray}
&&H_{1{\bf q}\omega} \nonumber \\
&&=A L_1 \frac{2\pi\omega^2}{\omega_p^4(cq)^2}
\omega^2(\omega^2-\omega_p^2) 
\left[\vert\sigma_{{\bf q}\omega}(t)\vert^2+\frac{1}{\omega^2}
\vert\dot{\sigma}_{{\bf q}\omega}(t)\vert^2\right] \nonumber \\
&&\equiv A L_1 C_{1{\bf q}\omega}\left[\vert \sigma_{{\bf q}\omega}(t)\vert^2+
\frac{1}{\omega^2}\vert\dot{\sigma}_{{\bf q}\omega}(t)\vert^2\right], 
\label{34}
\end{eqnarray}
and in region $\Lambda_2$
\begin{eqnarray}
&&H_{2{\bf q}\omega} \nonumber \\
&&= A L_2 \frac{2\pi\omega^2}{\omega_p^4(cq)^2}(\omega^2-\omega_p^2)^2 
\left[\vert\sigma_{{\bf q}\omega}(t)\vert^2+\frac{1}{\omega^2}
\vert\dot{\sigma}_{{\bf q}\omega}(t)\vert^2\right] \nonumber \\
&&\equiv A L_2 C_{2{\bf q}\omega}\left[\vert \sigma_{{\bf q}\omega}(t)\vert^2+ 
\frac{1}{\omega^2}\vert\dot{\sigma}_{{\bf q}\omega}(t)\vert^2\right]  . 
\label{35}
\end{eqnarray}

For obtaining the total contribution of these modes, of different frequencies 
$\omega$, to $\beta S(t,{\bf q})$, it would be necessary to sum on $\omega$
the quantity
\begin{equation} \label{36}
\beta S_{C\omega}(t,{\bf q})=\frac{1}{L_1 C_{1{\bf q}\omega}
+L_2 C_{2{\bf q}\omega}} f(\omega)\cos(\omega t).
\end{equation}
We were not able to perform this sum. 
Fortunately, for $\omega$ close to $\omega_p$, (\ref{35}) is negligible 
compared to (\ref{34}) and it will turn out that the sum on $\omega$ 
involves only such values of $\omega$ in the limit of small $q$. 
Thus, we can neglect the term $L_2 C_{2{\bf q}\omega}$ in (\ref{36}) writing
\begin{equation}
\beta S_{C\omega}(t,{\bf q})=\frac{1}{L_1C_{1{\bf q}\omega}}
f(\omega)\cos(\omega t), \label{37}
\end{equation}
and perform the sum on $\omega$ by replacing it by an integral 
like in Section IV. 
Therefore the contribution of these modes is
\begin{eqnarray}
S_C(t,{\bf q}) & = & \frac{1}{\pi}\int_{\omega_p^2+(cq)^2}^{\infty}
d(\omega^2) \nonumber \\ & & \times
\frac{\omega_p^4 f(\omega)\cos(\omega t)}{4\pi c\sqrt{\omega^2-\omega_p^2
-(cq)^2}\omega^4(\omega^2-\omega_p^2)} . \label{38}
\end{eqnarray}
After the change of variable $\omega^2=\omega_p^2+(cq)^2(1+u)$, (\ref{38})
becomes, in the limit $q\rightarrow 0$, 
\begin{eqnarray}
S_C(t,{\bf q}) & \sim & \frac{q}{4\pi^2}f(\omega_p)\cos(\omega_p t)
\int_0^{\infty} \frac{d u}{\sqrt{u}(1+u)} \nonumber \\
& = & \frac{q}{4\pi}f(\omega_p)\cos(\omega_p t). \label{39}
\end{eqnarray}

\section{Longitudinal modes}
In addition to the transverse modes studied up to now, in region $\Lambda_1$,
there are longitudinal modes of frequency $\omega_p$ (then 
$\epsilon(\omega_p)=0$) \cite{Jackson,LL}. 
They are solutions of the equations (\ref{4}-\ref{8}) with now $\rho\neq 0$ 
and ${\bf B}=0$; therefore they occur also in the non-retarded case already 
studied in \cite{J}. 
These longitudinal modes are a superposition of waves with the electric field 
parallel to the wavenumber vector. 
Here, we summarize the calculations of \cite{J}, using a slightly different 
notation. 

Since all modes have the same frequency but differ by the component $x$ of 
the wave vector, in (\ref{3}) $\sigma_{{\bf q}\omega}(t)$ must be replaced by
$\sigma_{{\bf q}k}(t)$. Since ${\bf B}=0$, (\ref{5}) reduces to
$\nabla\times{\bf E}=0$, and ${\bf E}$ is derivable from a potential. 
This potential is of the form 
\begin{equation} \label{40}
\phi=\left[a(t)\exp( i q_y y)+{\rm c.c.}\right]\sin(kx);
\end{equation}
this form ensures that the $x$ component of the electric field  
\begin{equation} \label{41}
E_{1x}=-k\left[a(t)\exp( i q_y y)+{\rm c.c.}\right]\cos(kx)
\end{equation}
is non-zero at $x=0$, thus is coupled by (\ref{15}) to $\sigma_{{\bf q}k}(t)$ 
(in region $\Lambda_2$, there are no longitudinal waves, thus ${\bf E}_2=0$). 
(\ref{15}) gives $-ka(t)=4\pi\sigma_{{\bf q}k}(t)$.
After $E_{2y}$ has been expressed from (\ref{40}), one finds for the energy 
\begin{equation} \label{42}
H_{{\bf q}k}=4\pi AL_1\frac{k^2+q^2}{k^2}\left[\vert \sigma_{{\bf q}k}(t)
\vert^2 + \frac{1}{\omega_p^2} \vert\dot{\sigma}_{{\bf q}k}(t)\vert^2\right] .
\end{equation}
After an integration on $k$, one finds for the contribution of the
longitudinal modes 
\begin{equation}
\beta S_D(t,{\bf q})=\frac{1}{2\pi^2}f(\omega_p)\cos(\omega_p t)
\int_0^K d k\, \frac{k^2}{k^2+q^2},    \label{43}
\end{equation}
where $K$ is a cut-off beyond which the use of collective variables breaks 
down. 
For small $q$, the integral in (\ref{43}) is $K-(\pi/2)q+O(q^2)$; therefore, 
$\beta S_D(t,{\bf q})$ has a kink singularity at ${\bf q=0}$ of the form  
\begin{equation} \label{44}
\beta S_D(t,{\bf q})\sim-\frac{q}{4\pi} f(\omega_p) \cos(\omega_p t).
\end{equation}

\section{Conclusion}
(\ref{39}) and (\ref{44}) cancel each other. The only contribution of order
$q$ to $\beta S(t,{\bf q})$ is (\ref{32}). Thus
\begin{equation} \label{45}
\beta S(t,{\bf q})\sim\frac{q}{4\pi},
\end{equation}
the classical non-retarded result at $t=0$, as announced in the Introduction.
This classical non-retarded result at $t=0$ was found previously as 
the result of two contributions: 
The one from the non-retarded surface plasmons $2q/(4\pi)$
and the one from the longitudinal bulk modes in the plasma $-q/(4\pi)$. 
Things are different for the present retarded result, which comes from 
the fields in vacuum, with a cancellation of the contributions of 
the transverse waves and the longitudinal waves in the plasma.

An open qualitative problem is to understand by a physical argument why the 
retarded quantum result at an arbitrary $t$ is so simple, not involving
$\hbar$ nor $t$. 

\medskip

\begin{acknowledgments}
L. \v{S}amaj is grateful to LPT for its very kind hospitality.
The support received from the European Science Foundation 
(ESF ``Methods of Integrable Systems, Geometry, Applied Mathematics'')
and from the Grant VEGA 2/6071/2008 is acknowledged. 
\end{acknowledgments}

\medskip

\appendix
*\section{}
The energy density of a one-component plasma (\ref{14}) can be obtained from
the Maxwell equations (\ref{4}-\ref{7}) and the force equation (\ref{8}).
Energy conservation can be expressed as \cite{Jackson} 
\begin{equation} \label{A.1}
\frac{1}{8\pi}\frac{\partial}{\partial t}(E^2+B^2)+{\bf J}\cdot{\bf E}=
-\nabla\cdot{\bf S},
\end{equation}
where 
\begin{equation} \label{A.2} 
{\bf S}=\frac{c}{4\pi}({\bf E}\times{\bf B})
\end{equation}
is the Poynting vector (energy flow) and ${\bf J}\cdot{\bf E}$ is the rate 
per unit volume of doing work by the electric field. (\ref{8}) gives 
\begin{equation} \label{A.3}
{\bf J}\cdot{\bf E}=\frac{2\pi}{\omega_p^2}\frac{\partial(J^2)}{\partial t}. 
\end{equation}
The lhs of (\ref{A.1}) is the time derivative of the energy density, 
thus which is (\ref{14}).

The last term of (\ref{14}) can be expressed in terms of ${\bf E}$. 
One has to be careful about the sign: since oscillatory terms are to be
discarded, $J^2=\vert {\bf J}\vert^2$ and $E^2=\vert {\bf E}\vert^2$. 
Using again (\ref{8}), at frequency $\omega$, one can replace the last term 
in (\ref{14}) by $\omega_p^2 E^2/(8\pi\omega^2)$. Taking into account 
(\ref{10}), we obtain
\begin{equation} \label{A.4}
U_1=\frac{1}{8\pi}\left[\frac{d(\omega\epsilon)}{d\omega}E^2+B^2\right],
\end{equation}
as written in \cite{LL}.

\bibliography{surface2}

\end{document}